\documentclass[aps,prb,reprint,asmath,amssymb,superscriptaddress,floatfix]{revtex4-2}
\usepackage{graphicx}
\usepackage[colorlinks, citecolor=blue, linkcolor=blue]{hyperref}
\usepackage{txfonts}

\newcommand{\FU}{Department of Physics and Dahlem Center for Complex Quantum Systems, Freie Universit{\"a}t Berlin, D-14195 Berlin, Germany}
\newcommand{\FNRS}{Fonds de la Recherche Scientifique, B-1000 Brussels, Belgium}
\newcommand{\NANOMAT}{Nanomat/Q-mat/CESAM, Universit{\'e} de Li{\`e}ge, B-4000 Sart Tilman, Belgium}
\newcommand{\IPCMS}{Institut de Physique et Chimie des Mat{\'e}riaux de Strasbourg, CNRS, Universit{\'e} de Strasbourg, 67034 Strasbourg, France}
\newcommand{\UCCS}{Department of Physics and Energy Science, University of Colorado at Colorado Springs, Colorado Springs, Colorado 80918, USA}
\newcommand{\CNN}{Centre de Nanosciences et de Nanotechnologies, CNRS, Universit{\'e} Paris-Saclay, 91120 Palaiseau, France}

\begin{document}

\title{Asymmetric skyrmion-antiskyrmion production in ultrathin ferromagnetic films}
\author{Ulrike Ritzmann}
\affiliation{\FU}
\author{Louise Desplat}
\affiliation{\IPCMS}
\author{Bertrand Dup{\'e}}
\affiliation{\FNRS}
\affiliation{\NANOMAT}
\author{Robert E. Camley}
\affiliation{\UCCS}
\author{Joo-Von Kim}
\email{joo-von.kim@c2n.upsaclay.fr}
\affiliation{\CNN}

\date{6 November 2020}

\begin{abstract}
Ultrathin ferromagnets with frustrated exchange and the Dzyaloshinskii-Moriya interaction can support topological solitons such as skyrmions and antiskyrmions, which are metastable and can be considered particle-antiparticle counterparts. When spin-orbit torques are applied, the motion of an isolated antiskyrmion driven beyond its Walker limit can generate skyrmion-antiskyrmion pairs. Here, we use atomistic spin dynamics simulations to shed light on the scattering processes involved in this pair generation. Under certain conditions a proliferation of these particles and antiparticles can appear with a growth rate and production asymmetry that depend on the strength of the chiral interactions and the dissipative component of the spin-orbit torques. These features are largely determined by scattering processes between antiskyrmions, which can be elastic or result in bound states or annihilation. \\
\\
DOI: \href{https://doi.org/10.1103/PhysRevB.102.174409}{10.1103/PhysRevB.102.174409} \hskip 5 mm Cite as U. Ritzmann \textit{et al}., Phys. Rev. B \textbf{102}, 174409 (2020).

\end{abstract}

\maketitle

%%%
%%	Introduction
%
\section{Introduction}

Condensed matter offers a fascinating test bed to explore different concepts in non-relativistic and relativistic quantum field theories~\cite{Coleman:2003ku}, with some prominent examples being massless Dirac quasiparticles in graphene~\cite{Novoselov:2005es}, Majorana fermions in superconductors~\cite{Fu:2008gu, Mourik:2012je, Rokhinson:2012ep}, and anyons in two-dimensional electron gases~\cite{Bartolomei:2020gs}. In ultrathin ferromagnets, chiral interactions of the Dzyaloshinskii-Moriya form~\cite{Dzyaloshinsky:1958vq, Moriya:1960go, Moriya:1960kc, Fert:1980hr, Crepieux:1998ux, Bogdanov:2001hr} allow for the existence of skyrmions~\cite{Bogdanov:1989vt, Bogdanov:1994bt}, which are topological soliton solutions to a nonlinear field theory bearing resemblance to a model for mesons and baryons proposed by Skyrme~\cite{Skyrme:1961vo, Skyrme:1962tr}. While these two-dimensional particles have been actively studied for their potential in information storage applications~\cite{Kiselev:2011cm, Sampaio:2013kn}, their original ties to nucleons have been revisited through three-dimensional extensions called hopfions~\cite{Sutcliffe:2017da}, which also provide an intriguing connection to Kelvin's proposal for a vortex theory of atoms~\cite{Thomson:1867}.

Pairs of skyrmions and antiskyrmions, their antiparticle counterpart, can be generated in a variety of ways, such as nucleation under local heating~\cite{Koshibae:2014fg}, homogeneous spin currents~\cite{Stier:2017ic, EverschorSitte:2018bn}, and surface acoustic waves~\cite{Yokouchi:2020cl}. Pairs also appear in ultrathin chiral ferromagnets with frustrated exchange interactions when the magnetization dynamics is driven by spin-orbit torques (SOTs)~\cite{Ritzmann:2018cc}. While both skyrmions and antiskyrmions are metastable states in such systems~\cite{Leonov:2015iz, Lin:2016hh, Rozsa:2017ii}, their motion can be qualitatively different under spin-orbit torques~\cite{Ritzmann:2018cc}. In particular, an antiskyrmion driven beyond its Walker limit can shed skyrmion-antiskyrmion pairs, much like the vortex-antivortex pairs produced during vortex core reversal~\cite{VanWaeyenberge:2006io}, which are then driven apart by the SOTs. Because such nonlinear processes are observed to involve a variety of creation and annihilation events involving particles and antiparticles, the intriguing analogies with high-energy physics compel us to explore whether this system could offer any insight, albeit tangential, into the more general question of matter-antimatter asymmetry in the universe. After all, the Sakharov conditions for baryogenesis~\cite{Sakharov:1967}, namely, baryon number violation, charge conjugation and combined charge conjugation-parity violation, and out-of-equilibrium interactions, appear to be naturally fulfilled in the aforementioned case: no conservation laws exist for the number of skyrmions and antiskyrmions, the Dzyaloshinskii-Moriya interaction (DMI) breaks chiral symmetry and lifts the degeneracy between skyrmion and antiskyrmions, and dissipative torques (spin-orbit and Gilbert damping) representing nonequilibrium processes play a crucial role in pair generation.

In this paper, we examine theoretically the microscopic processes leading to an imbalance in the number of skyrmions and antiskyrmions produced as a result of SOT-driven antiskyrmion dynamics. The remainder of this paper is organized as follows. In Sec. II, we describe the atomistic model used and the dynamics simulated. Section III discusses the main scattering processes that occur between an antiskyrmion and the generated skyrmion-antiskyrmion pair. Detailed phase diagrams of the generation processes are presented in Sec. IV, where the role of the SOTs and material parameters such as the strength of the Dzyaloshinskii-Moriya interaction and polarization angle are discussed. In Sec. V, we present the minimum-energy paths for two scattering processes. Finally, some discussion and concluding remarks are given in Sec. VI.

%%%
%	Model
%%%
\section{Model and method}
\begin{figure}%[hbt!]
\centering\includegraphics[width=8.5cm]{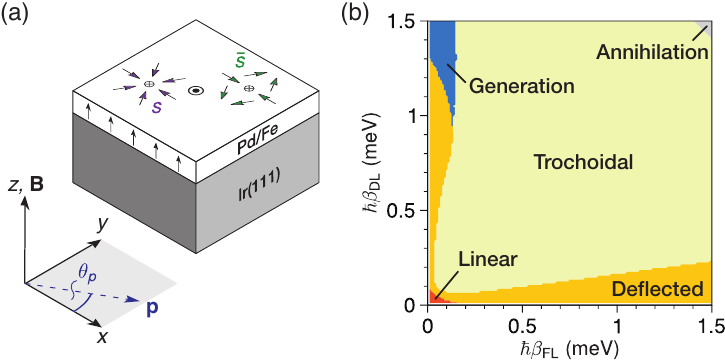}
\caption{(a) Film geometry illustrating the Pd/Fe bilayer on an Ir(111) substrate, with schematic illustrations of a skyrmion $s$ and antiskyrmion $\bar{s}$. $\mathbf{B}$ is the applied field and $\theta_p$ is the angle associated with the spin polarization vector $\mathbf{p}$. (b) Phase diagram of antiskyrmion dynamics under fieldlike (FL) and dampinglike (DL) spin-orbit torques~\cite{Ritzmann:2018cc}.
}
\label{fig:geometry}
\end{figure}

The system studied is illustrated in Fig.~\ref{fig:geometry}(a).
Following Refs.~\onlinecite{Romming:2013iq, Dupe:2014fc, Ritzmann:2018cc} we consider a ferromagnetic PdFe bilayer, which hosts the skyrmions $s$ and antiskyrmions $\bar{s}$, on an Ir(111) substrate through which we assume an electric current flows in the film plane, resulting in a spin current generated by the spin Hall effect flowing in the $z$ direction and polarized along $\mathbf{p}$, which is characterized by the angle $\theta_p$ measured from the $x$ axis. A magnetic field $\mathbf{B}$ is applied along the $z$ direction, which defines the uniform background state of the PdFe system. We model the magnetic properties of the PdFe film with a hexagonal lattice of magnetic moments $\mathbf{m}_i$, one atomic layer in thickness, whose dynamics is solved by time integration of the Landau-Lifshitz equation with Gilbert damping and spin-orbit torques,
\begin{eqnarray}
	\frac{d \mathbf{m} }{dt} = -\frac{1}{\hbar} \mathbf{m} \times \mathbf{B_{\mathrm{eff}}} + \alpha \mathbf{m} \times \frac{d \mathbf{m} }{dt} + \nonumber \\  \beta_\mathrm{FL} \mathbf{m} \times \mathbf{p}  + \beta_\mathrm{DL} \mathbf{m} \times \left( \mathbf{m} \times \mathbf{p} \right),
	\label{eq:LLG}
\end{eqnarray}
where $\alpha = 0.3$ is the damping constant and $\hbar \beta_\mathrm{FL}$ and $\hbar \beta_\mathrm{DL}$ characterize the strength of the fieldlike (FL) and dampinglike (DL) contributions of the spin-orbit torques, respectively, in meV. The effective field, $\mathbf{B}_i^{\mathrm{eff}}=-\partial \mathcal{H}/\partial \mathbf{m}_i$, is expressed here in units of energy and is derived from the Hamiltonian $\mathcal{H}$,
\begin{eqnarray}
\mathcal{H} = -\sum_{\langle ij \rangle} J_{ij} \mathbf{m}_i \cdot \mathbf{m}_j - \sum_{\langle ij \rangle} \mathbf{D}_{ij} \cdot \left(  \mathbf{m}_i \times \mathbf{m}_j \right) \nonumber \\ - \sum_{i} K m_{i,z}^2 - \sum_{i} \mathbf{B} \cdot \mu_\mathrm{s}\mathbf{m}_i.
\label{eq:Hamiltonian}	
\end{eqnarray}
The first term is the Heisenberg exchange interaction, which includes coupling up to ten nearest neighbors and involves frustrated exchange: $J_1 = 14.73$, $J_2=-1.95$, $J_3=-2.88$, $J_4=0.32$, $J_5=0.69$, $J_6=0.01$, $J_7=0.01$, $J_8=0.13$, $J_9=-0.14$, and $J_{10}=-0.28$, where all $J_{ij}$ are given in meV. The second term is the DMI between nearest neighbors, with $\mathbf{D}_{ij}$ oriented along $\hat{\mathbf{r}}_{ij} \times \hat{\mathbf{z}}$ and $ \| \mathbf{D}_{ij} \| = 1.0$ meV.  The third term describes a uniaxial anisotropy along the $z$ axis with $K = 0.7$ meV. The fourth term represents the Zeeman interaction with the applied field $\mathbf{B}$, where we take $\mu_s = 2.7\mu_\mathrm{B}$ for iron. The material parameters are obtained from first-principles calculations of the layered system in Fig.~\ref{fig:geometry}(a)~\cite{Dupe:2014fc}. We note that the applied field of 20 T is only slightly greater than the critical field $B_c$, $B=1.06 B_c$, below which the magnetic ground state comprises a skyrmion lattice phase. Under these conditions, \emph{both} isolated skyrmions and antiskyrmions are metastable states due to the frustrated exchange interactions, with skyrmions being energetically favored by the DMI.

Figure~\ref{fig:geometry}(b) represents the phase diagram, indicating different dynamical regimes under SOTs for a system in which the initial state comprises a single isolated antiskyrmion (the ``seed''). The linear, deflected, and trochoidal regimes denote the motion involving single-particle dynamics, while annihilation represents the region in which the seed loses its metastability. The focus here is on $s\bar{s}$ pair generation, which predominantly occurs under small fieldlike torques and large dampinglike torques. We simulated the dynamics in a variety of system sizes $L \times L$ with periodic boundary conditions, with $L$ ranging from 100 to 800 in order to mitigate finite-size effects that primarily involve collisions from generated particles re-entering the simulation area. The time integration of Eq.~(\ref{eq:LLG}) was performed using the Heun method with a time step of 1 fs.

\section{Scattering processes}
\begin{figure}%[hbt!]
\centering\includegraphics[width=8.5cm]{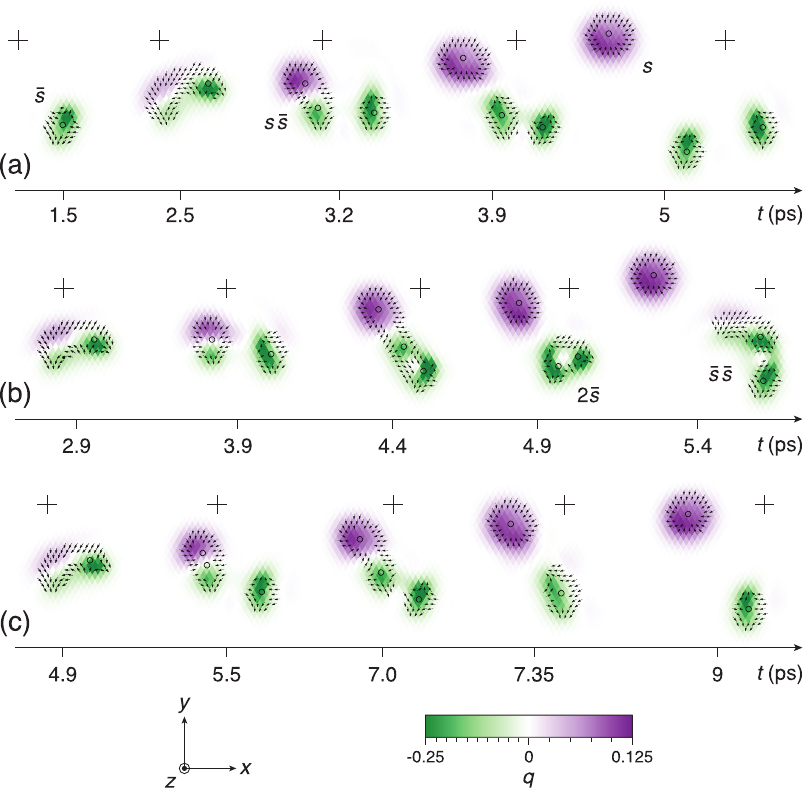}
\caption{Main scattering processes following pair generation from the seed $\bar{s}$ under SOT. (a) Maximal production, minimal asymmetry process $(N=2,\eta=0)$ leading to proliferation in which the generated $s\bar{s}$ pair splits and collision between the seed and generated $\bar{s}$ conserves skyrmion number. (b) $(N=2,\eta=0)$ process leading to premature saturation or stasis, where collision between the seed and generated $\bar{s}$ proceeds through a transient $Q=-2$ state ($2\bar{s}$) before decaying to an $\bar{s}\bar{s}$ bound pair, preventing further generation. (c) Minimal production, maximal asymmetry process ($N=1,\eta =1$) in which the generated $s\bar{s}$ pair splits and collision between the seed and generated $\bar{s}$ is inelastic, leading to annihilation of seed $\bar{s}$. Crosses denote the point of reference in the film plane and the color map indicates the charge density $q$ of a unit cell. Arrows are shown for moments for which $\sqrt{m_{i,x}^2+m_{i,y}^2} > 0.9$, and the open circles denote the approximate position of the core.}
\label{fig:pairprocess}
\end{figure}

The propensity for the initial seed $\bar{s}$ to produce particles ($s$) and antiparticles ($\bar{s}$) is determined by the scattering processes that immediately follow the formation of the $s\bar{s}$ pair, which depend on the strengths of $\beta_\mathrm{FL}$ and $\beta_\mathrm{DL}$. Three key scattering processes are illustrated in Fig.~\ref{fig:pairprocess} for $\theta_p = 0$.
The different processes illustrated typically occur for specific ranges of fieldlike and dampinglike parameters, as will be discussed later. We use a color map based on the local topological (skyrmion) charge density $q$, which is computed from three neighboring moments $\mathbf{m}_{i}, \mathbf{m}_{j}, \mathbf{m}_{k}$ as~\cite{Bottcher:2019hf}
\begin{equation}
q_{ijk} = -\frac{1}{2\pi} \tan^{-1}\left[ \frac{\mathbf{m}_{i} \cdot \left(\mathbf{m}_{j} \times \mathbf{m}_{k} \right)}{1+ \mathbf{m}_{i} \cdot \mathbf{m}_{j} + \mathbf{m}_{i}\cdot \mathbf{m}_{k} + \mathbf{m}_{j}\cdot \mathbf{m}_{k}}  \right].
\end{equation}
This represents the contribution from half a unit cell. We use $Q$ to denote the total charge, which represents a sum over $q_{ijk}$, and we adopt the convention where $Q=1$ for $s$ and $Q =-1$ for $\bar{s}$. The processes are characterized by their potential for particle production, measured by $N = N_s + N_{\bar{s}}$, which denotes the total numbers of skyrmions ($N_s$) and antiskyrmions ($N_{\bar{s}}$) produced from the initial antiskyrmion, and by the asymmetry in this production, which is measured by the parameter $\eta = (N_s - N_{\bar{s}})/N$. We consider only processes for which $N>0$ (the seed $\bar{s}$ is not included in this count). In Fig.~\ref{fig:pairprocess}(a) a maximal production ($N=2$), minimal asymmetry ($\eta=0$) scattering process leading to the proliferation of particles is shown for $(\hbar \beta_\mathrm{FL},\hbar \beta_\mathrm{DL}) = (0.02,1.5)$ meV. An $s\bar{s}$ pair nucleates from the tail of the $\bar{s}$ seed as it undergoes trochoidal motion, which then splits and is followed by a number-conserving collision between the two $\bar{s}$ particles. The $s$ particle escapes the zone of nucleation, and the two $\bar{s}$ particles become new sources of $s\bar{s}$ pair generation. In this scenario, $s$ and $\bar{s}$ are produced in equal numbers, and the process continues indefinitely but can be slowed by annihilation processes, which become more probable as the density of particles increases. In Fig.~\ref{fig:pairprocess}(b), a similar $N=2,\eta = 0$ process is shown for $(\hbar \beta_\mathrm{FL},\hbar \beta_\mathrm{DL}) = (0.1,1.35)$ meV, but here, the scattering between the two $\bar{s}$ results in a transient higher-order $Q=-2$ antiskyrmion state ($2\bar{s}$), which subsequently decays into an $\bar{s}\bar{s}$ bound pair that executes a rotational motion about its center of mass. As a result, further pair generation is suppressed. Figure~\ref{fig:pairprocess}(c) illustrates a minimal production ($N = 1$), maximal asymmetry ($\eta = 1$) process at $(\hbar \beta_\mathrm{FL},\hbar \beta_\mathrm{DL}) = (0.13,1.1)$ meV,  where the scattering between the seed and generated $\bar{s}$ results in a non-conservation process where the seed $\bar{s}$ is annihilated, which takes place via the creation and annihilation of a meron-antimeron~\cite{Desplat:2019dn}. This scattering event leaves the generated $s$ to propagate away and the surviving $\bar{s}$ to restart the process.

\begin{figure}
\centering\includegraphics[width=8.5cm]{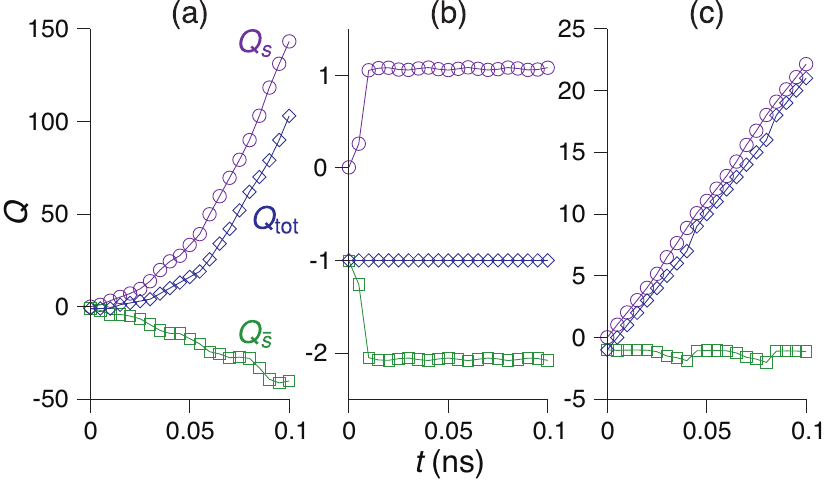}
\caption{Representative examples of different growth regimes of the total skyrmion charge, $Q$ for three values of $(\hbar \beta_\mathrm{FL},\hbar \beta_\mathrm{DL})$. (a) Proliferation, (0.02, 1.5) meV. (b) Stasis or premature saturation, (0.13, 1.1) meV. (c) Linear growth, (0.1, 1.35) meV.}
\label{fig:genrate}
\end{figure}

Examples of the growth rates are given in Fig.~\ref{fig:genrate}, where $Q(t)$ is shown for the three cases presented in Fig.~\ref{fig:pairprocess}.
The data are obtained from simulations of a $500 \times 500$ system over $0.1$ ns with $\theta_p = 0$. Above this timescale, propagating particles can reenter the simulation geometry through the periodic boundary conditions which result in spurious collisions and annihilation events. $Q_s$ and $Q_{\bar{s}}$ are found by summing over separately the contributions from $q_{ijk}>0$ and $q_{ijk}<0$, respectively. Figure~\ref{fig:genrate}(a) illustrates the growth where the process in Fig.~\ref{fig:pairprocess}(a) dominates, where a proliferation of particles can be seen. Unlike the single event in Fig.~\ref{fig:pairprocess}(a) the growth in Fig.~\ref{fig:genrate}(a) also comprises processes such as those described in Figs.~\ref{fig:pairprocess}(b) and ~\ref{fig:pairprocess}(c), which results in an overall asymmetry in the production and finite topological charge that increases with time. When the seed immediately undergoes the scattering process in Fig.~\ref{fig:pairprocess}(b), the generation stops for all future times, and a stasis regime is found [Fig.~\ref{fig:genrate}(b)]. Such processes can also occur after a certain time interval following proliferation, which results in premature saturation. Cases in which the scattering process in Fig.~\ref{fig:pairprocess}(c) repeats periodically results in an approximately linear growth in the number of skyrmions, which is shown in Fig.~\ref{fig:genrate}(c).

%%%
%	Phase diagrams
%%%
\section{Generation phase diagrams}
\begin{figure}%[hbt!]
\centering\includegraphics[width=8.5cm]{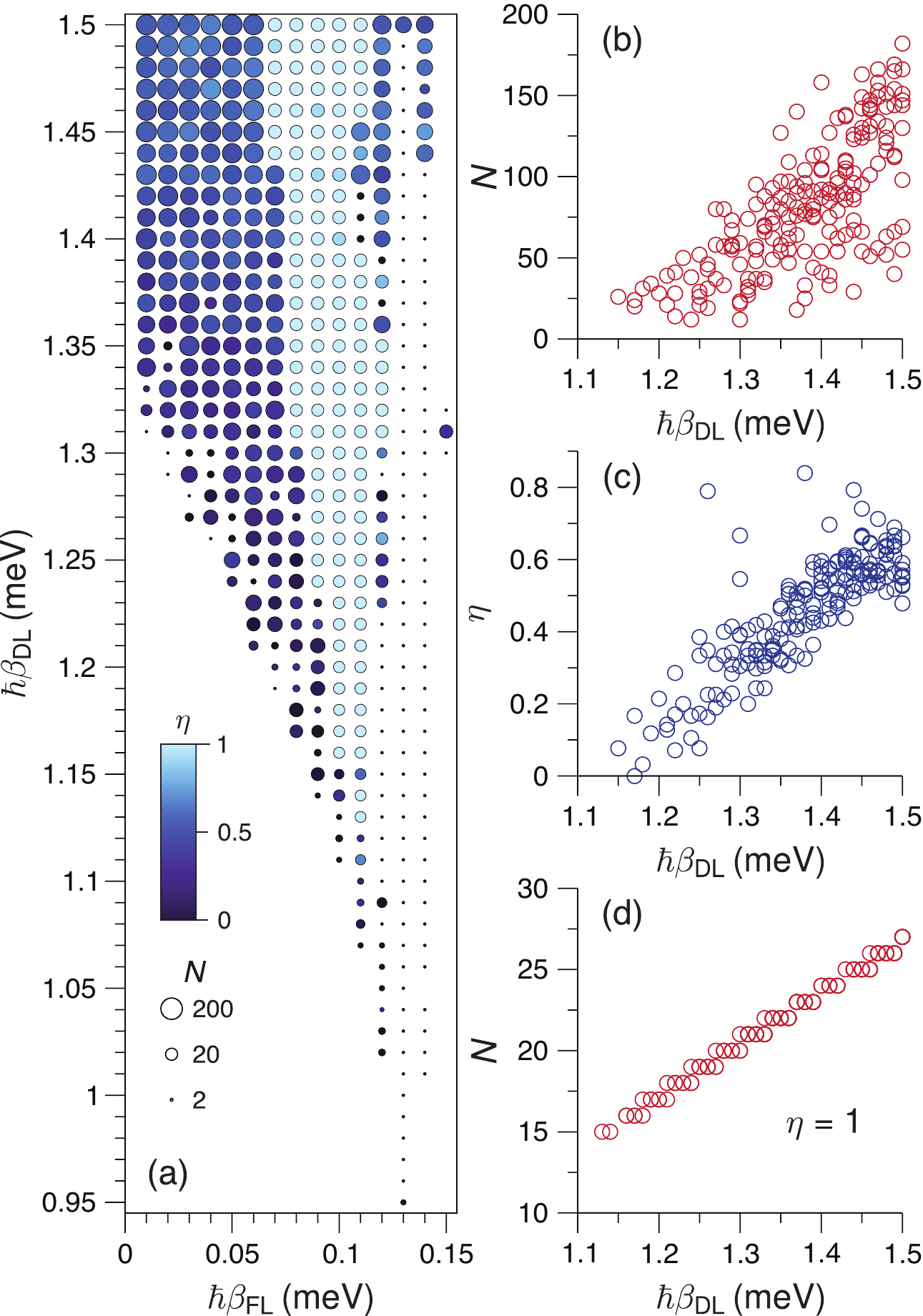}
\caption{(a) ($\beta_\mathrm{FL},\beta_\mathrm{DL}$) phase diagram illustrating the total number of skyrmions and antiskyrmions produced over 0.1 ns, where $N$ is represented by the circle size on a logarithmic scale and the asymmetry parameter is shown on a linear color scale. (b) $N$ and (c) $\eta$ as a function of DL torques for the proliferation regime (for different FL torques). (d) $N$ as a function of DL torques for linear growth ($\eta = 1$).}
\label{fig:phasediag}
\end{figure}

A ($\beta_\mathrm{FL},\beta_\mathrm{DL}$) phase diagram of the skyrmion production and asymmetry is presented in Fig.~\ref{fig:phasediag}(a).
As for Fig.~\ref{fig:genrate}, the data were obtained after simulating 0.1 ns on a $500 \times 500$ spin system with periodic boundary conditions and $\theta_p = 0$. The size of the circles represents $N$ on a logarithmic scale, while the color code represents $\eta$ on a linear scale.  Three different regimes can be identified visually as the strength of $\beta_\mathrm{FL}$ is increased. For low values of $\beta_\mathrm{FL}$ (primarily $\hbar\beta_\mathrm{FL} \lesssim 0.07$ meV), we observe a regime in which proliferation dominates where large numbers of $s$ and $\bar{s}$ are generated, which is mainly driven by the process in Fig.~\ref{fig:pairprocess}(a). Both $N$ and $\eta$ increase with the dampinglike torques in this regime, as shown in Figs.~\ref{fig:phasediag}(b) and \ref{fig:phasediag}(c), respectively, which can be understood from the fact that $\beta_\mathrm{DL}$ represents a nonconservative torque that transfers spin angular momentum into the system. For intermediate values of $\beta_\mathrm{FL}$ (primarily $0.08 \lesssim \hbar\beta_\mathrm{FL} \lesssim 0.11$ meV), a linear growth regime is seen which is characterized by $\eta \simeq 1$ and moderate values of $N$. As for the proliferation regime, the rate of production in the linear regime increases with $\beta_\mathrm{DL}$ as shown in Fig.~\ref{fig:phasediag}(d). Finally, for large values of $\beta_\mathrm{FL}$ (primarily $\hbar\beta_\mathrm{FL} \gtrsim 0.13$ meV) and close to the boundaries of the generation phase, we observe predominantly a stasis regime where generation stops after the nucleation of a single $s\bar{s}$ pair and the formation of a bound $\bar{s}\bar{s}$ state, as shown in Fig.~\ref{fig:pairprocess}(b).

\begin{figure}%[hbt!]
\centering\includegraphics[width=8.5cm]{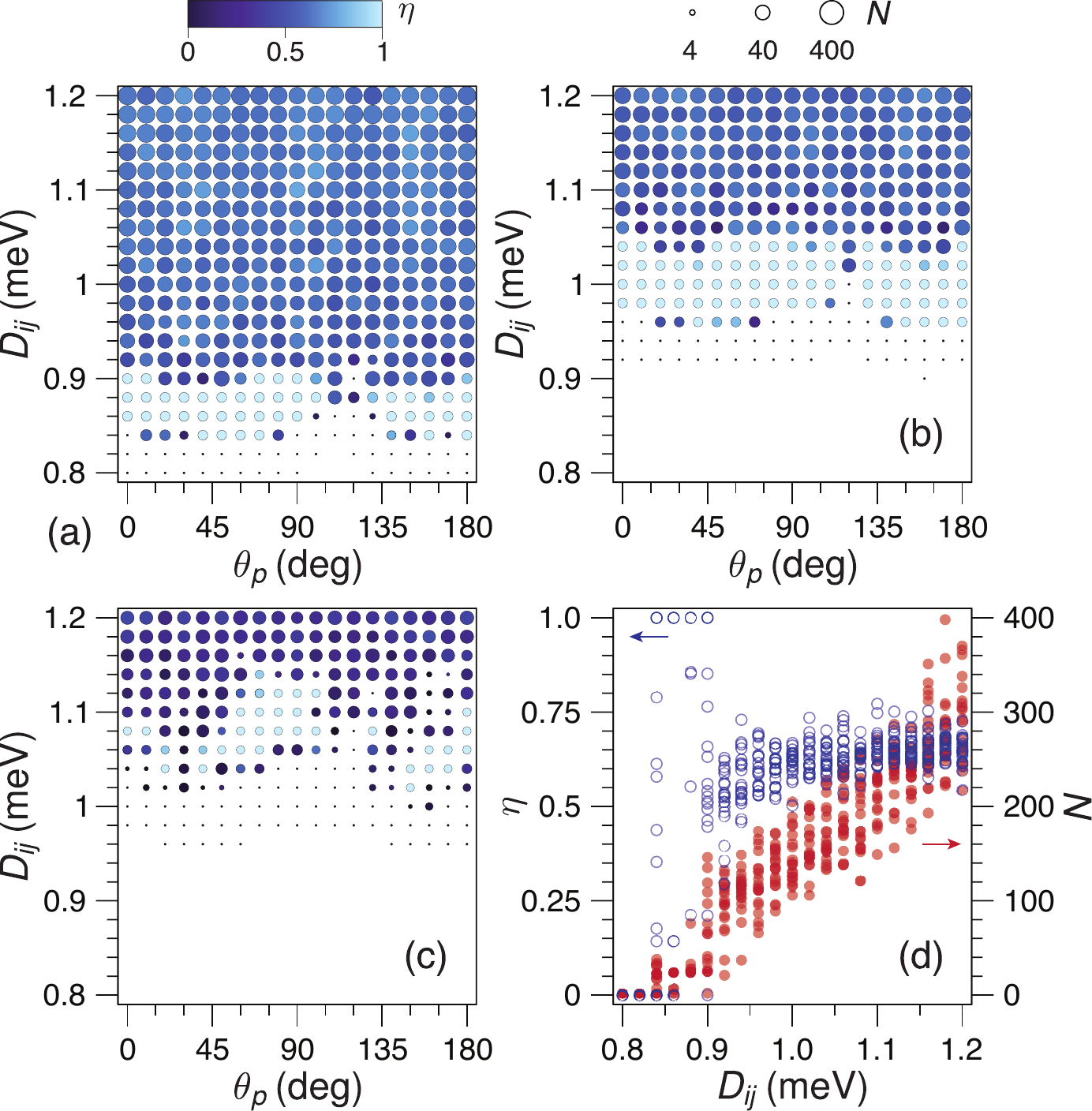}
\caption{($\theta_p,D_{ij}$) phase diagram illustrating the total number of skyrmions and antiskyrmions produced over 0.1 ns, where $N$ is represented by the circle size on a logarithmic scale and $\eta$ is shown on a linear color scale for (a) $(\beta_\mathrm{FL},\beta_\mathrm{DL}) =$ (0.02, 1.5) meV, (b) (0.1, 1.35) meV, and (c) (0.13, 1.1). (d) $\eta$ and $N$ as a function of $D_{ij}$ for the case in (a).}
\label{fig:thetad}
\end{figure}

The roles of DMI and the spin polarization angle are shown in Fig.~\ref{fig:thetad}, where $(\theta_p,D_{ij})$ phase diagrams for $N$ and $\eta$ are presented for the three distinct dynamical regimes discussed above: proliferation [(0.02, 1.5) meV, Fig.~\ref{fig:thetad}(a)], linear growth [(0.1, 1.35) meV, Fig.~\ref{fig:thetad}(b)], and stasis [(0.13, 1.1) meV, Fig.~\ref{fig:thetad}(c)], where the numbers in parentheses indicate values of $(\hbar \beta_\mathrm{FL},\hbar \beta_\mathrm{DL})$.
A weak dependence on $\theta_p$ can be seen. This arises from the interplay between the SOT-driven dynamics of the antiskyrmion helicity, which possesses twofold rotational symmetry about the antiskyrmion core in its rest state, and the underlying hexagonal lattice structure, which introduces a weak lattice potential that arises because of the compact nature of the core~\cite{Ritzmann:2018cc}. Variations in the magnitude of $D_{ij}$, on the other hand, lead to greater changes in the qualitative behavior, where transitions between stasis, linear growth, and proliferation can be seen as $D_{ij}$ is increased for all three base cases considered. This behavior is exemplified in Fig.~\ref{fig:thetad}(d), where $N$ and $\eta$ are shown as a function of $D_{ij}$ for the cases shown in Fig.~\ref{fig:thetad}(a). These results also suggest that a finite threshold for $D_{ij}$ is required for pair generation to take place, a threshold that is also dependent on the strength of the SOT applied.

%%%
%	MEPs
%%%
\section{Minimum-energy paths for merging and annihilation processes}
We can note that both stasis and proliferation states can be found at the phase boundaries. This results from the fact that the scattering processes in Figs.~\ref{fig:pairprocess}(b) and \ref{fig:pairprocess}(c) involve nearly identical energy barriers (in the absence of SOTs), where only slight differences in the relative helicities of the scattering $\bar{s}$ states determine the outcome. To see this, we look at minimum-energy paths (MEPs) on the multidimensional energy surface defined by the Hamiltonian in Eq.~(\ref{eq:Hamiltonian}) at $\beta_\mathrm{FL}=\beta_\mathrm{DL}=0$. We use the geodesic nudged elastic band method (GNEB)~\cite{Bessarab:2015method} to compute the MEPs, for which intermediate states of the system along the reaction coordinate are referred to as images. 
\begin{figure*}[hbt]
\centering\includegraphics[width=17.5cm]{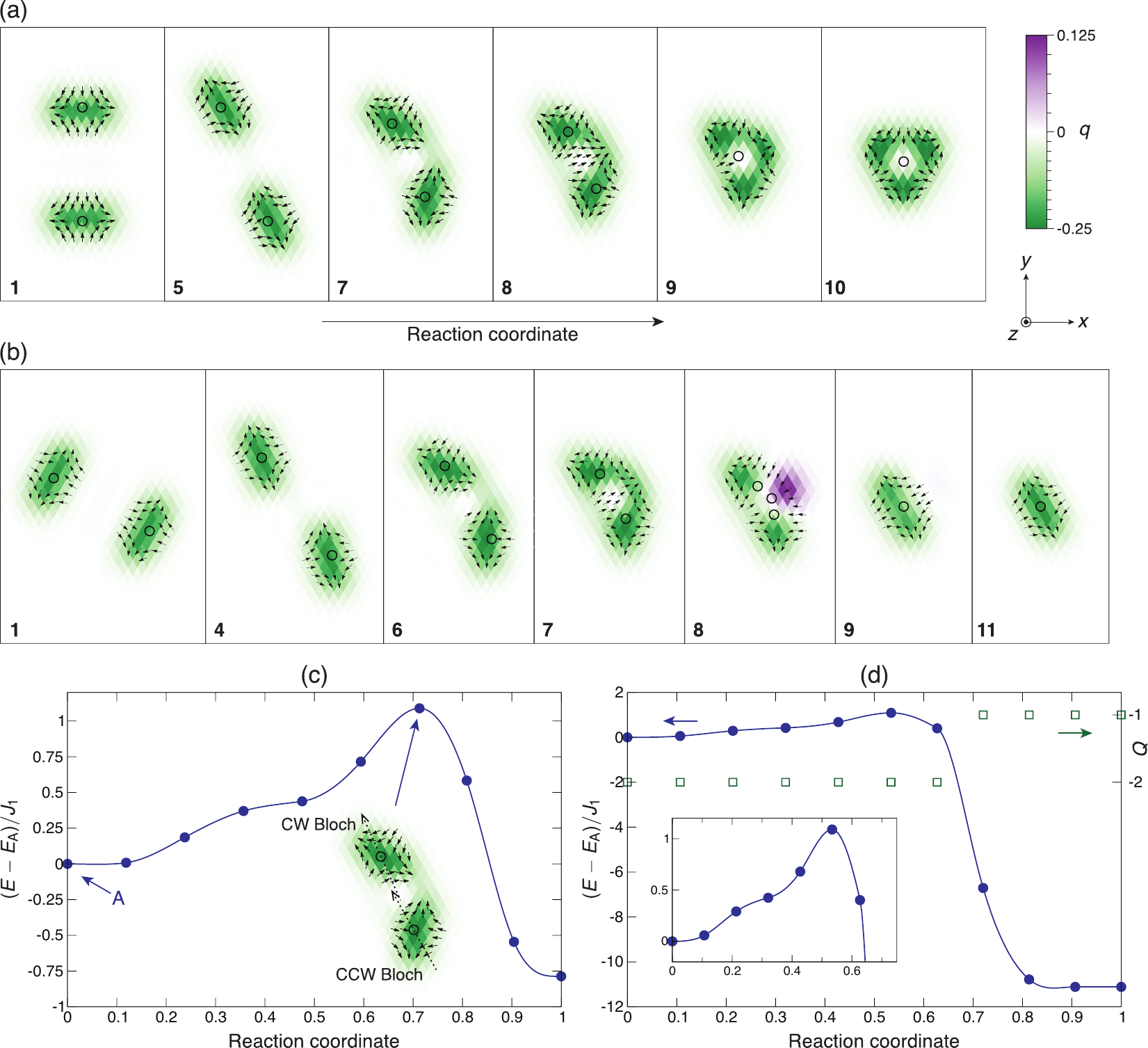}
\caption{(a, b) Minimum-energy paths for the merging of the $\bar{s}\bar{s}$ pair into (a) a $2\bar{s}$ state and (b) an $\bar{s}$ state. The image indices are given in the bottom left corner. The associated energy profile along the (normalized) reaction coordinate, where (c) corresponds to the paths that results in the $2\bar{s}$ state and (d) corresponds to the path that results in the $\bar{s}$ state. The total topological charge remains constant at $Q=-2$ in (c), while its variation with the reaction coordinate is shown in (d). The inset in (c) shows the saddle point configuration (image 7), where the dashed arrows indicate the reference axis along which the clockwise (CW) or counterclockwise (CCW) Bloch states are defined and through which the merging of $\bar{s}$ occurs. The inset in (d) represents an expanded view of the region around the energy barrier.}
\label{fig:meps}
\end{figure*}
First, the MEP for the merging into a higher-order $2\bar{s}$ state is shown in Fig.~\ref{fig:meps}(a), where the image index is shown in the bottom left corner.  The corresponding energy profile along the reaction coordinate is shown in Fig.~\ref{fig:meps}(c).  This path resembles the mechanism identified in Fig.~\ref{fig:pairprocess}(b), which, under SOTs, subsequently results in the formation of a bound $\bar{s}\bar{s}$ pair and suppresses generation. The initial state (A) in the GNEB method is set as a pair of metastable, isolated $\bar{s}$ states, where both $\bar{s}$ have the same helicity. The antiskyrmions then undergo a rotation of helicity, during which the total energy increases, to reach a higher-energy configuration at image 6. The next image, image 7, corresponds to the barrier top, in the form of a saddle point, and precedes the merging. At the saddle point, the antiskyrmions come into contact from the side and join through their counterclockwise and clockwise rotating Bloch axes, respectively, with a helicity difference of about $\pi$ rad. The corresponding energy barrier is found to be $\Delta E = 1.089 J_1$, where $J_1 = 14.73$ meV is the exchange constant for the Heisenberg interaction between nearest neighbors and is employed here as a characteristic energy scale. Subsequent images correspond to the merging into the final metastable $2\bar{s}$ state via the antiskyrmions' Bloch axes, accompanied by a decrease in the total energy of the system. The total topological charge remains constant throughout this process.

Next, we describe the path corresponding to the merging of the $\bar{s}\bar{s}$ pair into a single $\bar{s}$ via a process that does not conserve the total topological charge. The MEP is shown in Fig.~\ref{fig:meps}(b), with the corresponding energy profile shown in Fig.~\ref{fig:meps}(d). This mechanism resembles the process presented in Fig.~~\ref{fig:pairprocess}(c), through which an inelastic collision of two antiskyrmions results in the destruction of the seed and leads to a linear growth in the number of skyrmions [Fig.~\ref{fig:pairprocess}(c)]. Similar to the mechanism described above, the initial state is set as a pair of isolated, metastable $\bar{s}$ states, where both $\bar{s}$ have the same helicity. From there, the helicities of the antiskyrmions rotate as the energy increases, until the system reaches the barrier top at image 6. This state is very similar to the saddle point of the MEP in Fig.~\ref{fig:meps}(a), with, once more, a corresponding energy barrier of $\Delta E = 1.089 J_1$. However, the difference in the helicities seems to be, in this case, slightly inferior to $\pi$ rad. The following images correspond to the merging into a metastable single $\bar{s}$ state. This involves the destruction of one unit of negative topological charge, which occurs via the nucleation of a meron of charge $Q=+\frac{1}{2}$ at image 8. This is accompanied by a sharp decrease in the total energy of the system, as well as a drop in the total negative topological charge, from $-2$ to $-1$. The meron then annihilates with the extra antimeron of charge $Q=-\frac{1}{2}$, thus leaving a single $\bar{s}$ state of charge $Q=-1$ at image 9, accompanied by a further drop in the total energy.

The above results show that, in the generation regime, the scattering processes undergone by the $\bar{s}$ seed closely resemble the paths of minimum energy at zero SOT. Additionally, we find that the paths for the merging of the $\bar{s}\bar{s}$ pair into either a $2\bar{s}$ state or an $\bar{s}$ state traverse very similar saddle points, where only a small relative difference in the helicities appears to determine the fate of the final state. The associated energy barriers are practically identical and relatively low, of the order of $J_1$. This weak differentiation between the saddle points is in line with the fact that the boundaries of the phase diagram in Fig. 4 are not sharp and that small variations in the applied torques are sufficient to transition between the stasis and linear growth regimes.

%%%
%
%%%
\section{Discussion and concluding remarks}
With the frustrated exchange and in the absence of dipolar interactions, setting $D_{ij}$ to zero restores the chiral symmetry between skyrmions and antiskyrmions, where SOTs result in circular motion with opposite rotational sense for $s$ and $\bar{s}$~\cite{Leonov:2015iz, Lin:2016hh, Zhang:2017iv, Ritzmann:2018cc}. While the focus here has been on the consequences of generation from an antiskyrmion seed, the choice of an anisotropic form of the Dzyaloshinskii-Moriya interaction, i.e., one that energetically favors antiskyrmions over skyrmions~\cite{Nayak:2017hv, Hoffmann:2017kl, Camosi:2018eu,  Raeliarijaona:2018eg, Jena:2020db}, would result in the opposite behavior whereby skyrmion seeds would lead to pair generation and proliferation of antiskyrmions over skyrmions~\cite{Ritzmann:2018cc}.

Naturally, dipolar interactions are present in real materials, and their role has not been considered in this present study. This is justified for the following reasons. First, the long-range nature of dipolar interactions becomes apparent only as the film thickness is increased, i.e., beyond several nanometers. The system considered here is one atomic layer thick, which represents the limit in which the dipolar interaction is well described by a local approximation which results in the renormalization of the perpendicular magnetic anisotropy constant. Second, dipolar interactions favor a Bloch-like state for skyrmions and modify the energy dependence of the helicity for antiskyrmions. However, these corrections would almost be negligible in comparison with the strength of the frustrated exchange and Dzyaloshinskii-Moriya interactions considered. Finally, the inclusion of dipolar interactions would not suppress the Walker-like transition of the antiskyrmion dynamics, which results in pair generation.

In summary, we have presented results from atomistic spin dynamics simulations of skyrmion-antiskyrmion generation processes that result from the SOT-driven dynamics of an initial antiskyrmion state. Three fundamental scattering processes are identified, namely, elastic collisions, double-antiskyrmion bound states, and antiskyrmion annihilation, which form the basis of more complex generation processes leading to stasis, linear growth, and proliferation of particles. We investigated how the strength of the spin-orbit torques, including the orientation of the spin polarization with respect to the lattice, and the DMI constant affect the generation processes. Overall, the asymmetry in the production of particles and antiparticles from a given seed is driven by the strength of the chiral symmetry breaking, here measured by $D_{ij}$, and the nonequilibrium torques leading to pair generation, here characterized by $\beta_\mathrm{DL}$. Last, we investigated the paths of minimum energy at zero SOT for the two fundamental scattering processes that respectively lead to the stasis and linear growth regimes. We found that these resemble the processes undergone by the seed under SOT, and that the two mechanisms involve extremely similar saddle points, which explains the lack of sharp boundaries between the two regimes.

%%%
%	Acknowledgements
%%%
\begin{acknowledgments}
This work was supported by the Agence Nationale de la Recherche under Contract No. ANR-17-CE24-0025 (TOPSKY), the Deutsche Forschungsgemeinschaft via TRR 227, and the University of Strasbourg Institute for Advanced Study (USIAS) for via a fellowship, within the French national program ``Investment for the Future'' (IdEx-Unistra).
\end{acknowledgments}

\bibliography{articles}

\begin{thebibliography}{41}%
\makeatletter
\providecommand \@ifxundefined [1]{%
 \@ifx{#1\undefined}
}%
\providecommand \@ifnum [1]{%
 \ifnum #1\expandafter \@firstoftwo
 \else \expandafter \@secondoftwo
 \fi
}%
\providecommand \@ifx [1]{%
 \ifx #1\expandafter \@firstoftwo
 \else \expandafter \@secondoftwo
 \fi
}%
\providecommand \natexlab [1]{#1}%
\providecommand \enquote  [1]{``#1''}%
\providecommand \bibnamefont  [1]{#1}%
\providecommand \bibfnamefont [1]{#1}%
\providecommand \citenamefont [1]{#1}%
\providecommand \href@noop [0]{\@secondoftwo}%
\providecommand \href [0]{\begingroup \@sanitize@url \@href}%
\providecommand \@href[1]{\@@startlink{#1}\@@href}%
\providecommand \@@href[1]{\endgroup#1\@@endlink}%
\providecommand \@sanitize@url [0]{\catcode `\\12\catcode `\$12\catcode
  `\&12\catcode `\#12\catcode `\^12\catcode `\_12\catcode `\%12\relax}%
\providecommand \@@startlink[1]{}%
\providecommand \@@endlink[0]{}%
\providecommand \url  [0]{\begingroup\@sanitize@url \@url }%
\providecommand \@url [1]{\endgroup\@href {#1}{\urlprefix }}%
\providecommand \urlprefix  [0]{URL }%
\providecommand \Eprint [0]{\href }%
\providecommand \doibase [0]{https://doi.org/}%
\providecommand \selectlanguage [0]{\@gobble}%
\providecommand \bibinfo  [0]{\@secondoftwo}%
\providecommand \bibfield  [0]{\@secondoftwo}%
\providecommand \translation [1]{[#1]}%
\providecommand \BibitemOpen [0]{}%
\providecommand \bibitemStop [0]{}%
\providecommand \bibitemNoStop [0]{.\EOS\space}%
\providecommand \EOS [0]{\spacefactor3000\relax}%
\providecommand \BibitemShut  [1]{\csname bibitem#1\endcsname}%
\let\auto@bib@innerbib\@empty
%</preamble>
\bibitem [{\citenamefont {Coleman}(2003)}]{Coleman:2003ku}%
  \BibitemOpen
  \bibfield  {author} {\bibinfo {author} {\bibfnamefont {P.}~\bibnamefont
  {Coleman}},\ }\bibfield  {title} {\bibinfo {title} {{Many Body Physics:
  Unfinished Revolution}},\ }\href {https://doi.org/10.1007/s00023-003-0943-9}
  {\bibfield  {journal} {\bibinfo  {journal} {Annales Henri Poincar{\'e}}\
  }\textbf {\bibinfo {volume} {4}},\ \bibinfo {pages} {559} (\bibinfo {year}
  {2003})}\BibitemShut {NoStop}%
\bibitem [{\citenamefont {Novoselov}\ \emph {et~al.}(2005)\citenamefont
  {Novoselov}, \citenamefont {Geim}, \citenamefont {Morozov}, \citenamefont
  {Jiang}, \citenamefont {Katsnelson}, \citenamefont {Grigorieva},
  \citenamefont {Dubonos},\ and\ \citenamefont {Firsov}}]{Novoselov:2005es}%
  \BibitemOpen
  \bibfield  {author} {\bibinfo {author} {\bibfnamefont {K.~S.}\ \bibnamefont
  {Novoselov}}, \bibinfo {author} {\bibfnamefont {A.~K.}\ \bibnamefont {Geim}},
  \bibinfo {author} {\bibfnamefont {S.~V.}\ \bibnamefont {Morozov}}, \bibinfo
  {author} {\bibfnamefont {D.}~\bibnamefont {Jiang}}, \bibinfo {author}
  {\bibfnamefont {M.~I.}\ \bibnamefont {Katsnelson}}, \bibinfo {author}
  {\bibfnamefont {I.~V.}\ \bibnamefont {Grigorieva}}, \bibinfo {author}
  {\bibfnamefont {S.~V.}\ \bibnamefont {Dubonos}},\ and\ \bibinfo {author}
  {\bibfnamefont {A.~A.}\ \bibnamefont {Firsov}},\ }\bibfield  {title}
  {\bibinfo {title} {{Two-dimensional gas of massless Dirac fermions in
  graphene}},\ }\href {https://doi.org/Novoselov:2005es} {\bibfield  {journal}
  {\bibinfo  {journal} {Nature (London)}\ }\textbf {\bibinfo {volume} {438}},\
  \bibinfo {pages} {197} (\bibinfo {year} {2005})}\BibitemShut {NoStop}%
\bibitem [{\citenamefont {Fu}\ and\ \citenamefont {Kane}(2008)}]{Fu:2008gu}%
  \BibitemOpen
  \bibfield  {author} {\bibinfo {author} {\bibfnamefont {L.}~\bibnamefont
  {Fu}}\ and\ \bibinfo {author} {\bibfnamefont {C.~L.}\ \bibnamefont {Kane}},\
  }\bibfield  {title} {\bibinfo {title} {{Superconducting Proximity Effect and
  Majorana Fermions at the Surface of a Topological Insulator}},\ }\href
  {https://doi.org/10.1103/PhysRevLett.100.096407} {\bibfield  {journal}
  {\bibinfo  {journal} {Physical Review Letters}\ }\textbf {\bibinfo {volume}
  {100}},\ \bibinfo {pages} {096407} (\bibinfo {year} {2008})}\BibitemShut
  {NoStop}%
\bibitem [{\citenamefont {Mourik}\ \emph {et~al.}(2012)\citenamefont {Mourik},
  \citenamefont {Zuo}, \citenamefont {Frolov}, \citenamefont {Plissard},
  \citenamefont {Bakkers},\ and\ \citenamefont {Kouwenhoven}}]{Mourik:2012je}%
  \BibitemOpen
  \bibfield  {author} {\bibinfo {author} {\bibfnamefont {V.}~\bibnamefont
  {Mourik}}, \bibinfo {author} {\bibfnamefont {K.}~\bibnamefont {Zuo}},
  \bibinfo {author} {\bibfnamefont {S.~M.}\ \bibnamefont {Frolov}}, \bibinfo
  {author} {\bibfnamefont {S.~R.}\ \bibnamefont {Plissard}}, \bibinfo {author}
  {\bibfnamefont {E.~P. A.~M.}\ \bibnamefont {Bakkers}},\ and\ \bibinfo
  {author} {\bibfnamefont {L.~P.}\ \bibnamefont {Kouwenhoven}},\ }\bibfield
  {title} {\bibinfo {title} {{Signatures of Majorana Fermions in Hybrid
  Superconductor-Semiconductor Nanowire Devices}},\ }\href
  {https://doi.org/10.1126/science.1222360} {\bibfield  {journal} {\bibinfo
  {journal} {Science}\ }\textbf {\bibinfo {volume} {336}},\ \bibinfo {pages}
  {1003} (\bibinfo {year} {2012})}\BibitemShut {NoStop}%
\bibitem [{\citenamefont {Rokhinson}\ \emph {et~al.}(2012)\citenamefont
  {Rokhinson}, \citenamefont {Liu},\ and\ \citenamefont
  {Furdyna}}]{Rokhinson:2012ep}%
  \BibitemOpen
  \bibfield  {author} {\bibinfo {author} {\bibfnamefont {L.~P.}\ \bibnamefont
  {Rokhinson}}, \bibinfo {author} {\bibfnamefont {X.}~\bibnamefont {Liu}},\
  and\ \bibinfo {author} {\bibfnamefont {J.~K.}\ \bibnamefont {Furdyna}},\
  }\bibfield  {title} {\bibinfo {title} {{The fractional a.c. Josephson effect
  in a semiconductor--superconductor nanowire as a signature of Majorana
  particles}},\ }\href {https://doi.org/10.1038/nphys2429} {\bibfield
  {journal} {\bibinfo  {journal} {Nature Physics}\ }\textbf {\bibinfo {volume}
  {8}},\ \bibinfo {pages} {795} (\bibinfo {year} {2012})}\BibitemShut {NoStop}%
\bibitem [{\citenamefont {Bartolomei}\ \emph {et~al.}(2020)\citenamefont
  {Bartolomei}, \citenamefont {Kumar}, \citenamefont {Bisognin}, \citenamefont
  {Marguerite}, \citenamefont {Berroir}, \citenamefont {Bocquillon},
  \citenamefont {Pla{\c c}ais}, \citenamefont {Cavanna}, \citenamefont {Dong},
  \citenamefont {Gennser}, \citenamefont {Jin},\ and\ \citenamefont
  {F{\`e}ve}}]{Bartolomei:2020gs}%
  \BibitemOpen
  \bibfield  {author} {\bibinfo {author} {\bibfnamefont {H.}~\bibnamefont
  {Bartolomei}}, \bibinfo {author} {\bibfnamefont {M.}~\bibnamefont {Kumar}},
  \bibinfo {author} {\bibfnamefont {R.}~\bibnamefont {Bisognin}}, \bibinfo
  {author} {\bibfnamefont {A.}~\bibnamefont {Marguerite}}, \bibinfo {author}
  {\bibfnamefont {J.-M.}\ \bibnamefont {Berroir}}, \bibinfo {author}
  {\bibfnamefont {E.}~\bibnamefont {Bocquillon}}, \bibinfo {author}
  {\bibfnamefont {B.}~\bibnamefont {Pla{\c c}ais}}, \bibinfo {author}
  {\bibfnamefont {A.}~\bibnamefont {Cavanna}}, \bibinfo {author} {\bibfnamefont
  {Q.}~\bibnamefont {Dong}}, \bibinfo {author} {\bibfnamefont {U.}~\bibnamefont
  {Gennser}}, \bibinfo {author} {\bibfnamefont {Y.}~\bibnamefont {Jin}},\ and\
  \bibinfo {author} {\bibfnamefont {G.}~\bibnamefont {F{\`e}ve}},\ }\bibfield
  {title} {\bibinfo {title} {{Fractional statistics in anyon collisions.}},\
  }\href {https://doi.org/10.1126/science.aaz5601} {\bibfield  {journal}
  {\bibinfo  {journal} {Science}\ }\textbf {\bibinfo {volume} {368}},\ \bibinfo
  {pages} {173} (\bibinfo {year} {2020})}\BibitemShut {NoStop}%
\bibitem [{\citenamefont {Dzyaloshinsky}(1958)}]{Dzyaloshinsky:1958vq}%
  \BibitemOpen
  \bibfield  {author} {\bibinfo {author} {\bibfnamefont {I.}~\bibnamefont
  {Dzyaloshinsky}},\ }\bibfield  {title} {\bibinfo {title} {{A thermodynamic
  theory of "weak" ferromagnetism of antiferromagnetics}},\ }\href
  {https://doi.org/10.1016/0022-3697(58)90076-3} {\bibfield  {journal}
  {\bibinfo  {journal} {Journal of Physics and Chemistry of Solids}\ }\textbf
  {\bibinfo {volume} {4}},\ \bibinfo {pages} {241} (\bibinfo {year}
  {1958})}\BibitemShut {NoStop}%
\bibitem [{\citenamefont {Moriya}(1960{\natexlab{a}})}]{Moriya:1960go}%
  \BibitemOpen
  \bibfield  {author} {\bibinfo {author} {\bibfnamefont {T.}~\bibnamefont
  {Moriya}},\ }\bibfield  {title} {\bibinfo {title} {{Anisotropic Superexchange
  Interaction and Weak Ferromagnetism}},\ }\href
  {https://doi.org/10.1103/PhysRev.120.91} {\bibfield  {journal} {\bibinfo
  {journal} {Physical Review}\ }\textbf {\bibinfo {volume} {120}},\ \bibinfo
  {pages} {91} (\bibinfo {year} {1960}{\natexlab{a}})}\BibitemShut {NoStop}%
\bibitem [{\citenamefont {Moriya}(1960{\natexlab{b}})}]{Moriya:1960kc}%
  \BibitemOpen
  \bibfield  {author} {\bibinfo {author} {\bibfnamefont {T.}~\bibnamefont
  {Moriya}},\ }\bibfield  {title} {\bibinfo {title} {{New Mechanism of
  Anisotropic Superexchange Interaction}},\ }\href
  {https://doi.org/10.1103/PhysRevLett.4.228} {\bibfield  {journal} {\bibinfo
  {journal} {Physical Review Letters}\ }\textbf {\bibinfo {volume} {4}},\
  \bibinfo {pages} {228} (\bibinfo {year} {1960}{\natexlab{b}})}\BibitemShut
  {NoStop}%
\bibitem [{\citenamefont {Fert}\ and\ \citenamefont
  {Levy}(1980)}]{Fert:1980hr}%
  \BibitemOpen
  \bibfield  {author} {\bibinfo {author} {\bibfnamefont {A.}~\bibnamefont
  {Fert}}\ and\ \bibinfo {author} {\bibfnamefont {P.~M.}\ \bibnamefont
  {Levy}},\ }\bibfield  {title} {\bibinfo {title} {{Role of Anisotropic
  Exchange Interactions in Determining the Properties of Spin-Glasses}},\
  }\href {https://doi.org/10.1103/PhysRevLett.44.1538} {\bibfield  {journal}
  {\bibinfo  {journal} {Physical Review Letters}\ }\textbf {\bibinfo {volume}
  {44}},\ \bibinfo {pages} {1538} (\bibinfo {year} {1980})}\BibitemShut
  {NoStop}%
\bibitem [{\citenamefont {Cr{\'e}pieux}\ and\ \citenamefont
  {Lacroix}(1998)}]{Crepieux:1998ux}%
  \BibitemOpen
  \bibfield  {author} {\bibinfo {author} {\bibfnamefont {A.}~\bibnamefont
  {Cr{\'e}pieux}}\ and\ \bibinfo {author} {\bibfnamefont {C.}~\bibnamefont
  {Lacroix}},\ }\bibfield  {title} {\bibinfo {title}
  {{Dzyaloshinsky{\textemdash}Moriya interactions induced by symmetry breaking
  at a surface}},\ }\href {https://doi.org/10.1016/S0304-8853(97)01044-5}
  {\bibfield  {journal} {\bibinfo  {journal} {Journal of Magnetism and Magnetic
  Materials}\ }\textbf {\bibinfo {volume} {182}},\ \bibinfo {pages} {341}
  (\bibinfo {year} {1998})}\BibitemShut {NoStop}%
\bibitem [{\citenamefont {Bogdanov}\ and\ \citenamefont
  {R{\"o}{\ss}ler}(2001)}]{Bogdanov:2001hr}%
  \BibitemOpen
  \bibfield  {author} {\bibinfo {author} {\bibfnamefont {A.~N.}\ \bibnamefont
  {Bogdanov}}\ and\ \bibinfo {author} {\bibfnamefont {U.~K.}\ \bibnamefont
  {R{\"o}{\ss}ler}},\ }\bibfield  {title} {\bibinfo {title} {{Chiral Symmetry
  Breaking in Magnetic Thin Films and Multilayers}},\ }\href
  {https://doi.org/10.1103/PhysRevLett.87.037203} {\bibfield  {journal}
  {\bibinfo  {journal} {Physical Review Letters}\ }\textbf {\bibinfo {volume}
  {87}},\ \bibinfo {pages} {037203} (\bibinfo {year} {2001})}\BibitemShut
  {NoStop}%
\bibitem [{\citenamefont {Bogdanov}\ and\ \citenamefont
  {Yablonskii}(1989)}]{Bogdanov:1989vt}%
  \BibitemOpen
  \bibfield  {author} {\bibinfo {author} {\bibfnamefont {A.~N.}\ \bibnamefont
  {Bogdanov}}\ and\ \bibinfo {author} {\bibfnamefont {D.~A.}\ \bibnamefont
  {Yablonskii}},\ }\bibfield  {title} {\bibinfo {title} {{Thermodynamically
  stable ``vortices'' in magnetically ordered crystals. The mixed state of
  magnets}},\ }\href@noop {} {\bibfield  {journal} {\bibinfo  {journal}
  {Journal of Experimental and Theoretical Physics}\ }\textbf {\bibinfo
  {volume} {68}},\ \bibinfo {pages} {101} (\bibinfo {year} {1989})}\BibitemShut
  {NoStop}%
\bibitem [{\citenamefont {Bogdanov}\ and\ \citenamefont
  {Hubert}(1994)}]{Bogdanov:1994bt}%
  \BibitemOpen
  \bibfield  {author} {\bibinfo {author} {\bibfnamefont {A.}~\bibnamefont
  {Bogdanov}}\ and\ \bibinfo {author} {\bibfnamefont {A.}~\bibnamefont
  {Hubert}},\ }\bibfield  {title} {\bibinfo {title} {{Thermodynamically stable
  magnetic vortex states in magnetic crystals}},\ }\href
  {https://doi.org/10.1016/0304-8853(94)90046-9} {\bibfield  {journal}
  {\bibinfo  {journal} {Journal of Magnetism and Magnetic Materials}\ }\textbf
  {\bibinfo {volume} {138}},\ \bibinfo {pages} {255} (\bibinfo {year}
  {1994})}\BibitemShut {NoStop}%
\bibitem [{\citenamefont {Skyrme}(1961)}]{Skyrme:1961vo}%
  \BibitemOpen
  \bibfield  {author} {\bibinfo {author} {\bibfnamefont {T.~H.~R.}\
  \bibnamefont {Skyrme}},\ }\bibfield  {title} {\bibinfo {title} {{A Non-linear
  Field Theory}},\ }\href {https://doi.org/10.1098/rspa.1961.0018} {\bibfield
  {journal} {\bibinfo  {journal} {Proceedings of the Royal Society of London,
  Series A}\ }\textbf {\bibinfo {volume} {260}},\ \bibinfo {pages} {127}
  (\bibinfo {year} {1961})}\BibitemShut {NoStop}%
\bibitem [{\citenamefont {Skyrme}(1962)}]{Skyrme:1962tr}%
  \BibitemOpen
  \bibfield  {author} {\bibinfo {author} {\bibfnamefont {T.~H.~R.}\
  \bibnamefont {Skyrme}},\ }\bibfield  {title} {\bibinfo {title} {{A unified
  field theory of mesons and baryons}},\ }\href
  {https://doi.org/10.1016/0029-5582(62)90775-7} {\bibfield  {journal}
  {\bibinfo  {journal} {Nuclear Physics}\ }\textbf {\bibinfo {volume} {31}},\
  \bibinfo {pages} {556} (\bibinfo {year} {1962})}\BibitemShut {NoStop}%
\bibitem [{\citenamefont {Kiselev}\ \emph {et~al.}(2011)\citenamefont
  {Kiselev}, \citenamefont {Bogdanov}, \citenamefont {Sch{\"a}fer},\ and\
  \citenamefont {R{\"o}{\ss}ler}}]{Kiselev:2011cm}%
  \BibitemOpen
  \bibfield  {author} {\bibinfo {author} {\bibfnamefont {N.~S.}\ \bibnamefont
  {Kiselev}}, \bibinfo {author} {\bibfnamefont {A.~N.}\ \bibnamefont
  {Bogdanov}}, \bibinfo {author} {\bibfnamefont {R.}~\bibnamefont
  {Sch{\"a}fer}},\ and\ \bibinfo {author} {\bibfnamefont {U.~K.}\ \bibnamefont
  {R{\"o}{\ss}ler}},\ }\bibfield  {title} {\bibinfo {title} {{Chiral skyrmions
  in thin magnetic films: New objects for magnetic storage technologies?}},\
  }\href {https://doi.org/10.1088/0022-3727/44/39/392001} {\bibfield  {journal}
  {\bibinfo  {journal} {Journal of Physics D: Applied Physics}\ }\textbf
  {\bibinfo {volume} {44}},\ \bibinfo {pages} {392001} (\bibinfo {year}
  {2011})}\BibitemShut {NoStop}%
\bibitem [{\citenamefont {Sampaio}\ \emph {et~al.}(2013)\citenamefont
  {Sampaio}, \citenamefont {Cros}, \citenamefont {Rohart}, \citenamefont
  {Thiaville},\ and\ \citenamefont {Fert}}]{Sampaio:2013kn}%
  \BibitemOpen
  \bibfield  {author} {\bibinfo {author} {\bibfnamefont {J.}~\bibnamefont
  {Sampaio}}, \bibinfo {author} {\bibfnamefont {V.}~\bibnamefont {Cros}},
  \bibinfo {author} {\bibfnamefont {S.}~\bibnamefont {Rohart}}, \bibinfo
  {author} {\bibfnamefont {A.}~\bibnamefont {Thiaville}},\ and\ \bibinfo
  {author} {\bibfnamefont {A.}~\bibnamefont {Fert}},\ }\bibfield  {title}
  {\bibinfo {title} {{Nucleation, stability and current-induced motion of
  isolated magnetic skyrmions in nanostructures}},\ }\href
  {https://doi.org/10.1038/nnano.2013.210} {\bibfield  {journal} {\bibinfo
  {journal} {Nature Nanotechnology}\ }\textbf {\bibinfo {volume} {8}},\
  \bibinfo {pages} {839} (\bibinfo {year} {2013})}\BibitemShut {NoStop}%
\bibitem [{\citenamefont {Sutcliffe}(2017)}]{Sutcliffe:2017da}%
  \BibitemOpen
  \bibfield  {author} {\bibinfo {author} {\bibfnamefont {P.}~\bibnamefont
  {Sutcliffe}},\ }\bibfield  {title} {\bibinfo {title} {{Skyrmion Knots in
  Frustrated Magnets}},\ }\href
  {https://doi.org/10.1103/PhysRevLett.118.247203} {\bibfield  {journal}
  {\bibinfo  {journal} {Physical Review Letters}\ }\textbf {\bibinfo {volume}
  {118}},\ \bibinfo {pages} {247203} (\bibinfo {year} {2017})}\BibitemShut
  {NoStop}%
\bibitem [{\citenamefont {Thomson}(1867)}]{Thomson:1867}%
  \BibitemOpen
  \bibfield  {author} {\bibinfo {author} {\bibfnamefont {W.}~\bibnamefont
  {Thomson}},\ }\bibfield  {title} {\bibinfo {title} {{On Vortex Atoms}},\
  }\href@noop {} {\bibfield  {journal} {\bibinfo  {journal} {Proceedings of the
  Royal Society of Edinburgh}\ }\textbf {\bibinfo {volume} {6}},\ \bibinfo
  {pages} {94} (\bibinfo {year} {1867})}\BibitemShut {NoStop}%
\bibitem [{\citenamefont {Koshibae}\ and\ \citenamefont
  {Nagaosa}(2014)}]{Koshibae:2014fg}%
  \BibitemOpen
  \bibfield  {author} {\bibinfo {author} {\bibfnamefont {W.}~\bibnamefont
  {Koshibae}}\ and\ \bibinfo {author} {\bibfnamefont {N.}~\bibnamefont
  {Nagaosa}},\ }\bibfield  {title} {\bibinfo {title} {{Creation of skyrmions
  and antiskyrmions by local heating}},\ }\href
  {https://doi.org/Koshibae:2014fg} {\bibfield  {journal} {\bibinfo  {journal}
  {Nature Communications}\ }\textbf {\bibinfo {volume} {5}},\ \bibinfo {pages}
  {5148} (\bibinfo {year} {2014})}\BibitemShut {NoStop}%
\bibitem [{\citenamefont {Stier}\ \emph {et~al.}(2017)\citenamefont {Stier},
  \citenamefont {H{\"a}usler}, \citenamefont {Posske}, \citenamefont {Gurski},\
  and\ \citenamefont {Thorwart}}]{Stier:2017ic}%
  \BibitemOpen
  \bibfield  {author} {\bibinfo {author} {\bibfnamefont {M.}~\bibnamefont
  {Stier}}, \bibinfo {author} {\bibfnamefont {W.}~\bibnamefont {H{\"a}usler}},
  \bibinfo {author} {\bibfnamefont {T.}~\bibnamefont {Posske}}, \bibinfo
  {author} {\bibfnamefont {G.}~\bibnamefont {Gurski}},\ and\ \bibinfo {author}
  {\bibfnamefont {M.}~\bibnamefont {Thorwart}},\ }\bibfield  {title} {\bibinfo
  {title} {{Skyrmion{\textendash}Anti-Skyrmion Pair Creation by in-Plane
  Currents}},\ }\href {https://doi.org/10.1103/PhysRevLett.118.267203}
  {\bibfield  {journal} {\bibinfo  {journal} {Physical Review Letters}\
  }\textbf {\bibinfo {volume} {118}},\ \bibinfo {pages} {267203} (\bibinfo
  {year} {2017})}\BibitemShut {NoStop}%
\bibitem [{\citenamefont {Everschor-Sitte}\ \emph {et~al.}(2018)\citenamefont
  {Everschor-Sitte}, \citenamefont {Sitte}, \citenamefont {Valet},
  \citenamefont {Abanov},\ and\ \citenamefont
  {Sinova}}]{EverschorSitte:2018bn}%
  \BibitemOpen
  \bibfield  {author} {\bibinfo {author} {\bibfnamefont {K.}~\bibnamefont
  {Everschor-Sitte}}, \bibinfo {author} {\bibfnamefont {M.}~\bibnamefont
  {Sitte}}, \bibinfo {author} {\bibfnamefont {T.}~\bibnamefont {Valet}},
  \bibinfo {author} {\bibfnamefont {A.}~\bibnamefont {Abanov}},\ and\ \bibinfo
  {author} {\bibfnamefont {J.}~\bibnamefont {Sinova}},\ }\bibfield  {title}
  {\bibinfo {title} {{Skyrmion production on demand by homogeneous DC
  currents}},\ }\href {https://doi.org/10.1088/1367-2630/aa8569} {\bibfield
  {journal} {\bibinfo  {journal} {New Journal of Physics}\ }\textbf {\bibinfo
  {volume} {19}},\ \bibinfo {pages} {092001} (\bibinfo {year}
  {2018})}\BibitemShut {NoStop}%
\bibitem [{\citenamefont {Yokouchi}\ \emph {et~al.}(2020)\citenamefont
  {Yokouchi}, \citenamefont {Sugimoto}, \citenamefont {Rana}, \citenamefont
  {Seki}, \citenamefont {Ogawa}, \citenamefont {Kasai},\ and\ \citenamefont
  {Otani}}]{Yokouchi:2020cl}%
  \BibitemOpen
  \bibfield  {author} {\bibinfo {author} {\bibfnamefont {T.}~\bibnamefont
  {Yokouchi}}, \bibinfo {author} {\bibfnamefont {S.}~\bibnamefont {Sugimoto}},
  \bibinfo {author} {\bibfnamefont {B.}~\bibnamefont {Rana}}, \bibinfo {author}
  {\bibfnamefont {S.}~\bibnamefont {Seki}}, \bibinfo {author} {\bibfnamefont
  {N.}~\bibnamefont {Ogawa}}, \bibinfo {author} {\bibfnamefont
  {S.}~\bibnamefont {Kasai}},\ and\ \bibinfo {author} {\bibfnamefont
  {Y.}~\bibnamefont {Otani}},\ }\bibfield  {title} {\bibinfo {title} {{Creation
  of magnetic skyrmions by surface acoustic waves}},\ }\href
  {https://doi.org/10.1038/s41565-020-0661-1} {\bibfield  {journal} {\bibinfo
  {journal} {Nature Nanotechnology}\ }\textbf {\bibinfo {volume} {15}},\
  \bibinfo {pages} {361} (\bibinfo {year} {2020})}\BibitemShut {NoStop}%
\bibitem [{\citenamefont {Ritzmann}\ \emph {et~al.}(2018)\citenamefont
  {Ritzmann}, \citenamefont {von Malottki}, \citenamefont {Kim}, \citenamefont
  {Heinze}, \citenamefont {Sinova},\ and\ \citenamefont
  {Dup{\'e}}}]{Ritzmann:2018cc}%
  \BibitemOpen
  \bibfield  {author} {\bibinfo {author} {\bibfnamefont {U.}~\bibnamefont
  {Ritzmann}}, \bibinfo {author} {\bibfnamefont {S.}~\bibnamefont {von
  Malottki}}, \bibinfo {author} {\bibfnamefont {J.-V.}\ \bibnamefont {Kim}},
  \bibinfo {author} {\bibfnamefont {S.}~\bibnamefont {Heinze}}, \bibinfo
  {author} {\bibfnamefont {J.}~\bibnamefont {Sinova}},\ and\ \bibinfo {author}
  {\bibfnamefont {B.}~\bibnamefont {Dup{\'e}}},\ }\bibfield  {title} {\bibinfo
  {title} {{Trochoidal motion and pair generation in skyrmion and antiskyrmion
  dynamics under spin{\textendash}orbit torques}},\ }\href
  {https://doi.org/10.1038/s41928-018-0114-0} {\bibfield  {journal} {\bibinfo
  {journal} {Nature Electronics}\ }\textbf {\bibinfo {volume} {1}},\ \bibinfo
  {pages} {451} (\bibinfo {year} {2018})}\BibitemShut {NoStop}%
\bibitem [{\citenamefont {Leonov}\ and\ \citenamefont
  {Mostovoy}(2015)}]{Leonov:2015iz}%
  \BibitemOpen
  \bibfield  {author} {\bibinfo {author} {\bibfnamefont {A.~O.}\ \bibnamefont
  {Leonov}}\ and\ \bibinfo {author} {\bibfnamefont {M.}~\bibnamefont
  {Mostovoy}},\ }\bibfield  {title} {\bibinfo {title} {{Multiply periodic
  states and isolated skyrmions in an anisotropic frustrated magnet}},\ }\href
  {https://doi.org/10.1038/ncomms9275} {\bibfield  {journal} {\bibinfo
  {journal} {Nature Communications}\ }\textbf {\bibinfo {volume} {6}},\
  \bibinfo {pages} {8275} (\bibinfo {year} {2015})}\BibitemShut {NoStop}%
\bibitem [{\citenamefont {Lin}\ and\ \citenamefont
  {Hayami}(2016)}]{Lin:2016hh}%
  \BibitemOpen
  \bibfield  {author} {\bibinfo {author} {\bibfnamefont {S.-Z.}\ \bibnamefont
  {Lin}}\ and\ \bibinfo {author} {\bibfnamefont {S.}~\bibnamefont {Hayami}},\
  }\bibfield  {title} {\bibinfo {title} {{Ginzburg-Landau theory for skyrmions
  in inversion-symmetric magnets with competing interactions}},\ }\href
  {https://doi.org/10.1103/PhysRevB.93.064430} {\bibfield  {journal} {\bibinfo
  {journal} {Physical Review B}\ }\textbf {\bibinfo {volume} {93}},\ \bibinfo
  {pages} {064430} (\bibinfo {year} {2016})}\BibitemShut {NoStop}%
\bibitem [{\citenamefont {R{\'o}zsa}\ \emph {et~al.}(2017)\citenamefont
  {R{\'o}zsa}, \citenamefont {Palot{\'a}s}, \citenamefont {De{\'a}k},
  \citenamefont {Simon}, \citenamefont {Yanes}, \citenamefont {Udvardi},
  \citenamefont {Szunyogh},\ and\ \citenamefont {Nowak}}]{Rozsa:2017ii}%
  \BibitemOpen
  \bibfield  {author} {\bibinfo {author} {\bibfnamefont {L.}~\bibnamefont
  {R{\'o}zsa}}, \bibinfo {author} {\bibfnamefont {K.}~\bibnamefont
  {Palot{\'a}s}}, \bibinfo {author} {\bibfnamefont {A.}~\bibnamefont
  {De{\'a}k}}, \bibinfo {author} {\bibfnamefont {E.}~\bibnamefont {Simon}},
  \bibinfo {author} {\bibfnamefont {R.}~\bibnamefont {Yanes}}, \bibinfo
  {author} {\bibfnamefont {L.}~\bibnamefont {Udvardi}}, \bibinfo {author}
  {\bibfnamefont {L.}~\bibnamefont {Szunyogh}},\ and\ \bibinfo {author}
  {\bibfnamefont {U.}~\bibnamefont {Nowak}},\ }\bibfield  {title} {\bibinfo
  {title} {{Formation and stability of metastable skyrmionic spin structures
  with various topologies in an ultrathin film}},\ }\href
  {https://doi.org/10.1103/PhysRevB.95.094423} {\bibfield  {journal} {\bibinfo
  {journal} {Physical Review B}\ }\textbf {\bibinfo {volume} {95}},\ \bibinfo
  {pages} {094423} (\bibinfo {year} {2017})}\BibitemShut {NoStop}%
\bibitem [{\citenamefont {Van~Waeyenberge}\ \emph {et~al.}(2006)\citenamefont
  {Van~Waeyenberge}, \citenamefont {Puzic}, \citenamefont {Stoll},
  \citenamefont {Chou}, \citenamefont {Tyliszczak}, \citenamefont {Hertel},
  \citenamefont {F{\"a}hnle}, \citenamefont {Br{\"u}ckl}, \citenamefont {Rott},
  \citenamefont {Reiss}, \citenamefont {Neudecker}, \citenamefont {Weiss},
  \citenamefont {Back},\ and\ \citenamefont
  {Sch{\"u}tz}}]{VanWaeyenberge:2006io}%
  \BibitemOpen
  \bibfield  {author} {\bibinfo {author} {\bibfnamefont {B.}~\bibnamefont
  {Van~Waeyenberge}}, \bibinfo {author} {\bibfnamefont {A.}~\bibnamefont
  {Puzic}}, \bibinfo {author} {\bibfnamefont {H.}~\bibnamefont {Stoll}},
  \bibinfo {author} {\bibfnamefont {K.~W.}\ \bibnamefont {Chou}}, \bibinfo
  {author} {\bibfnamefont {T.}~\bibnamefont {Tyliszczak}}, \bibinfo {author}
  {\bibfnamefont {R.}~\bibnamefont {Hertel}}, \bibinfo {author} {\bibfnamefont
  {M.}~\bibnamefont {F{\"a}hnle}}, \bibinfo {author} {\bibfnamefont
  {H.}~\bibnamefont {Br{\"u}ckl}}, \bibinfo {author} {\bibfnamefont
  {K.}~\bibnamefont {Rott}}, \bibinfo {author} {\bibfnamefont {G.}~\bibnamefont
  {Reiss}}, \bibinfo {author} {\bibfnamefont {I.}~\bibnamefont {Neudecker}},
  \bibinfo {author} {\bibfnamefont {D.}~\bibnamefont {Weiss}}, \bibinfo
  {author} {\bibfnamefont {C.~H.}\ \bibnamefont {Back}},\ and\ \bibinfo
  {author} {\bibfnamefont {G.}~\bibnamefont {Sch{\"u}tz}},\ }\bibfield  {title}
  {\bibinfo {title} {{Magnetic vortex core reversal by excitation with short
  bursts of an alternating field}},\ }\href
  {https://doi.org/10.1038/nature05240} {\bibfield  {journal} {\bibinfo
  {journal} {Nature (London)}\ }\textbf {\bibinfo {volume} {444}},\ \bibinfo
  {pages} {461} (\bibinfo {year} {2006})}\BibitemShut {NoStop}%
\bibitem [{\citenamefont {Sakharov}(1991)}]{Sakharov:1967}%
  \BibitemOpen
  \bibfield  {author} {\bibinfo {author} {\bibfnamefont {A.~D.}\ \bibnamefont
  {Sakharov}},\ }\bibfield  {title} {\bibinfo {title} {{Violation of CP
  invariance, C asymmetry, and baryon asymmetry of the universe}},\ }\href
  {https://doi.org/10.1070/PU1991v034n05ABEH002497} {\bibfield  {journal}
  {\bibinfo  {journal} {Soviet Physics Uspekhi}\ }\textbf {\bibinfo {volume}
  {34}},\ \bibinfo {pages} {392} (\bibinfo {year} {1991})}\BibitemShut
  {NoStop}%
\bibitem [{\citenamefont {Romming}\ \emph {et~al.}(2013)\citenamefont
  {Romming}, \citenamefont {Hanneken}, \citenamefont {Menzel}, \citenamefont
  {Bickel}, \citenamefont {Wolter}, \citenamefont {Von~Bergmann}, \citenamefont
  {Kubetzka},\ and\ \citenamefont {Wiesendanger}}]{Romming:2013iq}%
  \BibitemOpen
  \bibfield  {author} {\bibinfo {author} {\bibfnamefont {N.}~\bibnamefont
  {Romming}}, \bibinfo {author} {\bibfnamefont {C.}~\bibnamefont {Hanneken}},
  \bibinfo {author} {\bibfnamefont {M.}~\bibnamefont {Menzel}}, \bibinfo
  {author} {\bibfnamefont {J.~E.}\ \bibnamefont {Bickel}}, \bibinfo {author}
  {\bibfnamefont {B.}~\bibnamefont {Wolter}}, \bibinfo {author} {\bibfnamefont
  {K.}~\bibnamefont {Von~Bergmann}}, \bibinfo {author} {\bibfnamefont
  {A.}~\bibnamefont {Kubetzka}},\ and\ \bibinfo {author} {\bibfnamefont
  {R.}~\bibnamefont {Wiesendanger}},\ }\bibfield  {title} {\bibinfo {title}
  {{Writing and Deleting Single Magnetic Skyrmions}},\ }\href
  {https://doi.org/10.1126/science.1240573} {\bibfield  {journal} {\bibinfo
  {journal} {Science}\ }\textbf {\bibinfo {volume} {341}},\ \bibinfo {pages}
  {636} (\bibinfo {year} {2013})}\BibitemShut {NoStop}%
\bibitem [{\citenamefont {Dup{\'e}}\ \emph {et~al.}(2014)\citenamefont
  {Dup{\'e}}, \citenamefont {Hoffmann}, \citenamefont {Paillard},\ and\
  \citenamefont {Heinze}}]{Dupe:2014fc}%
  \BibitemOpen
  \bibfield  {author} {\bibinfo {author} {\bibfnamefont {B.}~\bibnamefont
  {Dup{\'e}}}, \bibinfo {author} {\bibfnamefont {M.}~\bibnamefont {Hoffmann}},
  \bibinfo {author} {\bibfnamefont {C.}~\bibnamefont {Paillard}},\ and\
  \bibinfo {author} {\bibfnamefont {S.}~\bibnamefont {Heinze}},\ }\bibfield
  {title} {\bibinfo {title} {{Tailoring magnetic skyrmions in ultra-thin
  transition metal films}},\ }\href {https://doi.org/10.1038/ncomms5030}
  {\bibfield  {journal} {\bibinfo  {journal} {Nature Communications}\ }\textbf
  {\bibinfo {volume} {5}},\ \bibinfo {pages} {4030} (\bibinfo {year}
  {2014})}\BibitemShut {NoStop}%
\bibitem [{\citenamefont {B{\"o}ttcher}\ \emph {et~al.}(2019)\citenamefont
  {B{\"o}ttcher}, \citenamefont {Heinze}, \citenamefont {Egorov}, \citenamefont
  {Sinova},\ and\ \citenamefont {Dup{\'e}}}]{Bottcher:2019hf}%
  \BibitemOpen
  \bibfield  {author} {\bibinfo {author} {\bibfnamefont {M.}~\bibnamefont
  {B{\"o}ttcher}}, \bibinfo {author} {\bibfnamefont {S.}~\bibnamefont
  {Heinze}}, \bibinfo {author} {\bibfnamefont {S.}~\bibnamefont {Egorov}},
  \bibinfo {author} {\bibfnamefont {J.}~\bibnamefont {Sinova}},\ and\ \bibinfo
  {author} {\bibfnamefont {B.}~\bibnamefont {Dup{\'e}}},\ }\bibfield  {title}
  {\bibinfo {title} {{B{\textendash}T phase diagram of Pd/Fe/Ir(111) computed
  with parallel tempering Monte Carlo}},\ }\href
  {https://doi.org/10.1088/1367-2630/aae282} {\bibfield  {journal} {\bibinfo
  {journal} {New Journal of Physics}\ }\textbf {\bibinfo {volume} {20}},\
  \bibinfo {pages} {103014} (\bibinfo {year} {2019})}\BibitemShut {NoStop}%
\bibitem [{\citenamefont {Desplat}\ \emph {et~al.}(2019)\citenamefont
  {Desplat}, \citenamefont {Kim},\ and\ \citenamefont
  {Stamps}}]{Desplat:2019dn}%
  \BibitemOpen
  \bibfield  {author} {\bibinfo {author} {\bibfnamefont {L.}~\bibnamefont
  {Desplat}}, \bibinfo {author} {\bibfnamefont {J.-V.}\ \bibnamefont {Kim}},\
  and\ \bibinfo {author} {\bibfnamefont {R.~L.}\ \bibnamefont {Stamps}},\
  }\bibfield  {title} {\bibinfo {title} {{Paths to annihilation of first- and
  second-order (anti)skyrmions via (anti)meron nucleation on the frustrated
  square lattice}},\ }\href {https://doi.org/10.1103/PhysRevB.99.174409}
  {\bibfield  {journal} {\bibinfo  {journal} {Physical Review B}\ }\textbf
  {\bibinfo {volume} {99}},\ \bibinfo {pages} {174409} (\bibinfo {year}
  {2019})}\BibitemShut {NoStop}%
\bibitem [{\citenamefont {Bessarab}\ \emph {et~al.}(2015)\citenamefont
  {Bessarab}, \citenamefont {Uzdin},\ and\ \citenamefont
  {Jonsson}}]{Bessarab:2015method}%
  \BibitemOpen
  \bibfield  {author} {\bibinfo {author} {\bibfnamefont {P.~F.}\ \bibnamefont
  {Bessarab}}, \bibinfo {author} {\bibfnamefont {V.~M.}\ \bibnamefont
  {Uzdin}},\ and\ \bibinfo {author} {\bibfnamefont {H.}~\bibnamefont
  {Jonsson}},\ }\bibfield  {title} {\bibinfo {title} {{Method for finding
  mechanism and activation energy of magnetic transitions, applied to skyrmions
  and antivortex annihilation}},\ }\href
  {https://doi.org/10.1016/j.cpc.2015.07.001} {\bibfield  {journal} {\bibinfo
  {journal} {Computer Physics Communications}\ }\textbf {\bibinfo {volume}
  {196}},\ \bibinfo {pages} {335} (\bibinfo {year} {2015})}\BibitemShut
  {NoStop}%
\bibitem [{\citenamefont {Zhang}\ \emph {et~al.}(2017)\citenamefont {Zhang},
  \citenamefont {Xia}, \citenamefont {Zhou}, \citenamefont {Liu}, \citenamefont
  {Zhang},\ and\ \citenamefont {Ezawa}}]{Zhang:2017iv}%
  \BibitemOpen
  \bibfield  {author} {\bibinfo {author} {\bibfnamefont {X.}~\bibnamefont
  {Zhang}}, \bibinfo {author} {\bibfnamefont {J.}~\bibnamefont {Xia}}, \bibinfo
  {author} {\bibfnamefont {Y.}~\bibnamefont {Zhou}}, \bibinfo {author}
  {\bibfnamefont {X.}~\bibnamefont {Liu}}, \bibinfo {author} {\bibfnamefont
  {H.}~\bibnamefont {Zhang}},\ and\ \bibinfo {author} {\bibfnamefont
  {M.}~\bibnamefont {Ezawa}},\ }\bibfield  {title} {\bibinfo {title} {{Skyrmion
  dynamics~in a frustrated ferromagnetic film~and current-induced helicity
  locking-unlocking transition.}},\ }\href
  {https://doi.org/10.1038/s41467-017-01785-w} {\bibfield  {journal} {\bibinfo
  {journal} {Nature Communications}\ }\textbf {\bibinfo {volume} {8}},\
  \bibinfo {pages} {1717} (\bibinfo {year} {2017})}\BibitemShut {NoStop}%
\bibitem [{\citenamefont {Nayak}\ \emph {et~al.}(2017)\citenamefont {Nayak},
  \citenamefont {Kumar}, \citenamefont {Ma}, \citenamefont {Werner},
  \citenamefont {Pippel}, \citenamefont {Sahoo}, \citenamefont {Damay},
  \citenamefont {R{\"o}{\ss}ler}, \citenamefont {Felser},\ and\ \citenamefont
  {Parkin}}]{Nayak:2017hv}%
  \BibitemOpen
  \bibfield  {author} {\bibinfo {author} {\bibfnamefont {A.~K.}\ \bibnamefont
  {Nayak}}, \bibinfo {author} {\bibfnamefont {V.}~\bibnamefont {Kumar}},
  \bibinfo {author} {\bibfnamefont {T.}~\bibnamefont {Ma}}, \bibinfo {author}
  {\bibfnamefont {P.}~\bibnamefont {Werner}}, \bibinfo {author} {\bibfnamefont
  {E.}~\bibnamefont {Pippel}}, \bibinfo {author} {\bibfnamefont
  {R.}~\bibnamefont {Sahoo}}, \bibinfo {author} {\bibfnamefont
  {F.}~\bibnamefont {Damay}}, \bibinfo {author} {\bibfnamefont {U.~K.}\
  \bibnamefont {R{\"o}{\ss}ler}}, \bibinfo {author} {\bibfnamefont
  {C.}~\bibnamefont {Felser}},\ and\ \bibinfo {author} {\bibfnamefont
  {S.~S.~P.}\ \bibnamefont {Parkin}},\ }\bibfield  {title} {\bibinfo {title}
  {{Magnetic antiskyrmions above room temperature in tetragonal Heusler
  materials}},\ }\href {https://doi.org/10.1038/nature23466} {\bibfield
  {journal} {\bibinfo  {journal} {Nature (London)}\ }\textbf {\bibinfo {volume}
  {548}},\ \bibinfo {pages} {561} (\bibinfo {year} {2017})}\BibitemShut
  {NoStop}%
\bibitem [{\citenamefont {Hoffmann}\ \emph {et~al.}(2017)\citenamefont
  {Hoffmann}, \citenamefont {Zimmermann}, \citenamefont {M{\"u}ller},
  \citenamefont {Sch{\"u}rhoff}, \citenamefont {Kiselev}, \citenamefont
  {Melcher},\ and\ \citenamefont {Bl{\"u}gel}}]{Hoffmann:2017kl}%
  \BibitemOpen
  \bibfield  {author} {\bibinfo {author} {\bibfnamefont {M.}~\bibnamefont
  {Hoffmann}}, \bibinfo {author} {\bibfnamefont {B.}~\bibnamefont
  {Zimmermann}}, \bibinfo {author} {\bibfnamefont {G.~P.}\ \bibnamefont
  {M{\"u}ller}}, \bibinfo {author} {\bibfnamefont {D.}~\bibnamefont
  {Sch{\"u}rhoff}}, \bibinfo {author} {\bibfnamefont {N.~S.}\ \bibnamefont
  {Kiselev}}, \bibinfo {author} {\bibfnamefont {C.}~\bibnamefont {Melcher}},\
  and\ \bibinfo {author} {\bibfnamefont {S.}~\bibnamefont {Bl{\"u}gel}},\
  }\bibfield  {title} {\bibinfo {title} {{Antiskyrmions stabilized at
  interfaces by anisotropic Dzyaloshinskii-Moriya interactions}},\ }\href
  {https://doi.org/10.1038/s41467-017-00313-0} {\bibfield  {journal} {\bibinfo
  {journal} {Nature Communications}\ }\textbf {\bibinfo {volume} {8}},\
  \bibinfo {pages} {308} (\bibinfo {year} {2017})}\BibitemShut {NoStop}%
\bibitem [{\citenamefont {Camosi}\ \emph {et~al.}(2018)\citenamefont {Camosi},
  \citenamefont {Rougemaille}, \citenamefont {Fruchart}, \citenamefont
  {Vogel},\ and\ \citenamefont {Rohart}}]{Camosi:2018eu}%
  \BibitemOpen
  \bibfield  {author} {\bibinfo {author} {\bibfnamefont {L.}~\bibnamefont
  {Camosi}}, \bibinfo {author} {\bibfnamefont {N.}~\bibnamefont {Rougemaille}},
  \bibinfo {author} {\bibfnamefont {O.}~\bibnamefont {Fruchart}}, \bibinfo
  {author} {\bibfnamefont {J.}~\bibnamefont {Vogel}},\ and\ \bibinfo {author}
  {\bibfnamefont {S.}~\bibnamefont {Rohart}},\ }\bibfield  {title} {\bibinfo
  {title} {{Micromagnetics of antiskyrmions in ultrathin films}},\ }\href
  {https://doi.org/10.1103/PhysRevB.97.134404} {\bibfield  {journal} {\bibinfo
  {journal} {Physical Review B}\ }\textbf {\bibinfo {volume} {97}},\ \bibinfo
  {pages} {134404} (\bibinfo {year} {2018})}\BibitemShut {NoStop}%
\bibitem [{\citenamefont {Raeliarijaona}\ \emph {et~al.}(2018)\citenamefont
  {Raeliarijaona}, \citenamefont {Nepal},\ and\ \citenamefont
  {Kovalev}}]{Raeliarijaona:2018eg}%
  \BibitemOpen
  \bibfield  {author} {\bibinfo {author} {\bibfnamefont {A.}~\bibnamefont
  {Raeliarijaona}}, \bibinfo {author} {\bibfnamefont {R.}~\bibnamefont
  {Nepal}},\ and\ \bibinfo {author} {\bibfnamefont {A.~A.}\ \bibnamefont
  {Kovalev}},\ }\bibfield  {title} {\bibinfo {title} {{Boundary twists,
  instabilities, and creation of skyrmions and antiskyrmions}},\ }\href
  {https://doi.org/10.1103/PhysRevMaterials.2.124401} {\bibfield  {journal}
  {\bibinfo  {journal} {Physical Review Materials}\ }\textbf {\bibinfo {volume}
  {2}},\ \bibinfo {pages} {124401} (\bibinfo {year} {2018})}\BibitemShut
  {NoStop}%
\bibitem [{\citenamefont {Jena}\ \emph {et~al.}(2020)\citenamefont {Jena},
  \citenamefont {G{\"o}bel}, \citenamefont {Ma}, \citenamefont {Kumar},
  \citenamefont {Saha}, \citenamefont {Mertig}, \citenamefont {Felser},\ and\
  \citenamefont {Parkin}}]{Jena:2020db}%
  \BibitemOpen
  \bibfield  {author} {\bibinfo {author} {\bibfnamefont {J.}~\bibnamefont
  {Jena}}, \bibinfo {author} {\bibfnamefont {B.}~\bibnamefont {G{\"o}bel}},
  \bibinfo {author} {\bibfnamefont {T.}~\bibnamefont {Ma}}, \bibinfo {author}
  {\bibfnamefont {V.}~\bibnamefont {Kumar}}, \bibinfo {author} {\bibfnamefont
  {R.}~\bibnamefont {Saha}}, \bibinfo {author} {\bibfnamefont {I.}~\bibnamefont
  {Mertig}}, \bibinfo {author} {\bibfnamefont {C.}~\bibnamefont {Felser}},\
  and\ \bibinfo {author} {\bibfnamefont {S.~S.~P.}\ \bibnamefont {Parkin}},\
  }\bibfield  {title} {\bibinfo {title} {{Elliptical Bloch skyrmion chiral
  twins in an antiskyrmion system.}},\ }\href
  {https://doi.org/10.1038/s41467-020-14925-6} {\bibfield  {journal} {\bibinfo
  {journal} {Nature Communications}\ }\textbf {\bibinfo {volume} {11}},\
  \bibinfo {pages} {1115} (\bibinfo {year} {2020})}\BibitemShut {NoStop}%
\end{thebibliography}%
\end{document}